\documentclass{aastex}                 
\usepackage{spr-astr-addons}          
\usepackage{natbib}
\usepackage{epstopdf}
\usepackage{graphicx}
\usepackage{bm}
\usepackage[breaklinks=true,colorlinks=true,linkcolor=blue,urlcolor=blue,citecolor=blue]{hyperref}

\hypersetup{colorlinks,citecolor=blue,linkcolor=blue,urlcolor=blue}

\def\la{\;\raise0.3ex\hbox{$<$\kern-0.75em\raise-1.1ex\hbox{$\sim$}}\;}
\def\ga{\;\raise0.3ex\hbox{$>$\kern-0.75em\raise-1.1ex\hbox{$\sim$}}\;}

\begin{document}

%

\title{Power-spectrum simulations of radial redshift distributions}

\shorttitle{Power-spectrum simulations}

\shortauthors{Ryabinkov,  Kaminker}

\author{A. I. Ryabinkov}
\and
\author{A. D. Kaminker}  


\altaffiltext{}{Ioffe Institute, Politekhnicheskaya
	26, 194021,         \\ 
     St Petersburg, Russia  \\
ryabin60@gmail.com,\,   kam.astro@mail.ioffe.ru }

\begin{abstract}
On the base of the simplest model of a
modulation
of 3D Gaussian field in  $k$-space  
we produce a set of
simulations to bring out the
effects of a modulating
function $f_{\rm mod} (k)=f_1 (k) + f_2 (k)$
on power spectra of radial (shell-like)
distributions of cosmological objects, where
a model function $f_1 (k)$
reproduces the smoothed power spectrum
of underlying 3D density fluctuations,
while  $f_2 (k)$ is  
a wiggling function imitating
the baryon acoustic oscillations (BAO).
It is shown that some
excess of realizations
of simulated  radial distributions
actually displays quasi-periodical
components  with periods 
about a characteristic scale
$2\pi/k \sim  100~h^{-1}$~Mpc
detected as power-spectrum peaks  
in vicinity of the first maximum
of the modulation function  $f_2 (k)$.
We revised our previous estimations of 
the significance of such peaks
and found that they were largely
overestimated. Thereby
quasi-periodical components  
appearing in some radial distributions of matter
are likely to be
stochastic (rather than determinative), while
the  amplitudes  of the respective spectral peaks
can be quite noticeable.
They are partly enhanced by smooth
part of the modulating function $f_1(k)$
and,  to a far lesser extent,  by effects of the BAO 
(i.e. $f_2(k)$).
The results of the simulations match quite well
with statistical properties of the radial
distributions of the brightest cluster galaxies (BCGs).
\end{abstract}


%
\section{Introduction}
\label{sec:intro}

In this paper we consider statistical 
properties of the radial 
(shell-like) distributions of matter, 
that is a set of cosmological objects 
(characterized by redshifts)
observed in various directions 
in dependence only on their comoving line-of-sight distances.
We focus on a study of possible quasi-periodical 
components in the radial distribution of matter 
and  estimations of their significance.
The study of radial distribution of matter inhomogeneities
is of special interest to cosmology because its results allow
twofold interpretations, either  temporal or spatial ones. 
In particular, the temporal interpretation dominated 
in literature over long time
in the context of possible incorporation of wave processes
in the cosmological evolution
(e.g., \citealt{mor91, arkosh08, harhir08,  hirkom10}). 
The spatial interpretation of quasi-periodicities in the 
radial distributions occurs much less frequently
because it assumes some fraction of spatial order 
at least in some parts of the Universe 
(e.g.,  \citealt{einast97a, einast97b, einast16}).

One can treat our statistical approach 
as a somewhat different view on well-known statistics.
So in our previous papers  
(e.g.,   \citealt{rkk13, rk14},\ 
hereafter Paper~I and II)
we discussed an appearance of 
quasi-periodical components in the radial distributions  
of a few samples of redshifted objects 
adhering  to the  spatial interpretation.
The reason was that  the main scale 
revealed in these works 
$\sim 100~h^{-1}$~Mpc,
where $h=H_0/100$~km~s$^{-1}$~Mpc$^{-1}$, 
$H_0$ -- the present Hubble constant, 
was a spatial scale 
approximately corresponding to the  
Baryon Acoustic Oscillations (BAO)  (e.g., 
\citealt{eh98, eht98,  esw07, bh10,  percetal10,   
anderetal12,  kazetal14, anderetal14, alam17, ross17, varg-mag18}        
and references therein). 

It is well-known that 
the phenomenon of the BAO is the oscillatory dependence of
the power spectrum of luminous-matter fluctuations on a
wave number $k$. 
Amplitudes  at some  $k$ in the 
Fourier space  were enhanced
and at  others -- reduced during the recombination epoch.
These enhancements and suppressions 
manifest itself 
as quasi-periodical variations imprinted in the power spectrum
with the period  $2 \pi /r_s$  corresponding to so called
standard ``ruler''  $r_s$.
Such oscillations might  be superimposed on
random fluctuations of density (and velocity)
of luminous matter in the subsequent epochs 
(e.g., \citealt{sz70, pyu70,  zn83})
and  in principle  could promote 
the emergence of secondary quasi-periodicities
in the radial distribution of the objects. 

We proceeded from an assumption that
the radial distributions of cosmological objects  
may exhibit peaks  in their power spectra 
due  either to stochastic processes or as  footprints 
of actual periodical components. 
Here we evaluate  significance of such
peaks visible in the radial power spectra 
and estimate probability for the peaks to  occur 
in vicinity of the BAO scale. 
We consider the simplest model 
of the Gaussian field modulation in the $k$-space
accounting for two factors:
(i) a smooth galaxy power spectrum (e.g., \citealt{fkp94})
described by a function $f_1(k)$,     
(ii) damped oscillations of the power spectrum  
described by  a model function  $f_2(k)$, which    
is designed with using the approximation of 
\cite{bg03} to simulate the BAO effect.
Our main  simplifications are concerned 
with  full disregard 
of evolutionary effects and a
consideration of  pure spatial  
coordinate systems (CS)  --- Cartesian or  
spherical ones ---  conventionally treated in 
the present time. It makes sense to note 
that just the evolution can lead to  
non-linearity and non-Gaussian effects
which we do not consider here. 
We neglect also a velocity field of galaxies.

All spatial coordinates $x, y, z$ as well as coordinates
$k_x, k_y, k_z$ in the Fourier space  are measured 
in cosmological units $h^{-1}$~Mpc and $h$~Mpc$^{-1}$,
respectively, 
keeping in  mind possible  applications
of the results to a flat cosmological (comoving) space
(e.g.,   Sect.~\ref{sec:bcg}).  

Actually, we proceed from the Gaussian mock
data in real (comoving) space,  produce  
a modulated random field
in the $k$-space  and then
transform  the field back to
the real space.   This allows us
to examine
an occurrence of quasi-periodicities
in the radial distributions  and  infer
a conclusion about 
a probability to find those or other  
spatial periodic components. 
We  consider such components 
as {\it quasi}-periodical
because of limited intervals of line-of-sight distances
used in all analyzed samples    
and, as it is shown below, 
because of  the fact that
an appearance of such 
components has stochastic
nature in principle  and should be described  
in terms of probabilities. 

Our main results are that the BAO modulation
can not provide  
relatively high amplitudes of the peaks in the radial
power spectra declared in Papers~I and II.
Thereby the hypothesis about an impact of the BAO 
is not confirmed. 
On the other hand, it is shown that the high
amplitudes can be induced by basic fluctuations
of 3D power spectrum, which
could be treated as a {\it cosmic variance}. 
That strongly reduces 
significance of the peaks. Note that our new  evaluations
and our conclusions may be
referred  to both spatial and temporal interpretation.
    
In Sect.~\ref{sec:bd} we determine  
basic values and definitions used in our simulations. 
In Sect.~\ref{sec:ms} 
we introduce a modulating function  $f_{\rm mod}(k)$
and present results of the statistical analysis
of simulated  radial  distributions; in 
Sect.~\ref{sec:bcg}  we compare the results  with those
obtained in a similar way for samples of 
spectroscopic redshifts $z$ detected for 
so-called brightest 
cluster galaxies  (BCGs) or the most luminous galaxies 
in the clusters. Conclusions and discussions 
of the results are  given in Sect.~\ref{sec:cd}.

\section{BASIC  DEFINITIONS}
\label{sec:bd}  
We start our simulations in the Cartesian CS and generate
homogeneous, isotropic  Gaussian mock fields $G(x, y, z)$ 
as discrete cubic matrices with ${\cal N}_{bin}^3$ cells (bins) and 
the mean M$[G]$ equal to the variance D$[G]$. For certainty we choose 
M$[G]=$D$[G]=100$. 
 
As the second step we produce a Fourier transform into ${\bf k}$-space 
for each realization of $G(x, y, z)$ :
\begin{equation}
F_G ({\bf k}) = {1 \over  L^3}\, \int \, {\rm d}^3{\bf r}\, 
                        G({\bf r})\, e^{-i{\bf k}{\bf r}},
\label{FG}
\end{equation}
where $L={\cal N}_{bin}  \Delta_b$ is a linear size of a cubic 
box under investigation, 
$\Delta_b$ is a size of a cell (bin), 
in this
work we put $\Delta_b = 10~h^{-1}$~Mpc. 
Components of ${\bf k}$ in 
the Cartesian CS -- $k_x, k_y, k_z$ form also cubic matrices with
the same number of bins ${\cal N}_{bin}^3$.  
As a result of Eq.~(\ref{FG}) 
we obtain two 3D-matrices corresponding to real Re$(F_G)$ and imaginary
Im$(F_G)$  parts of the complex  values $F_G$. 
To return back to the real space one can produce 
an inverse Fourier transform  
with inclusion of a modulating function 
$f_{\rm mod}(k)$
\begin{equation}
U({\bf r})={L^3 \over  (2\pi)^3} \, \int \, {\rm d}^3 {\bf k}\, F_G({\bf k})\,
                            \sqrt{f_{\rm mod}(k)}\, e^{i{\bf k}{\bf r}},
\label{U}
\end{equation}
where $k=\sqrt{k_x^2 + k_y^2 + k_z^2}$,\  
$U({\bf r})=U(x, y, z)$ is
modulated random field  
which is an original subject of the further
statistical simulations.

Following standard definition 
one can introduce a variance of the  modulated field $U ({\bf r})$:
\begin{equation}
{\rm D} [U] = \frac{\sum_i\ \sum_j\ \sum_l\,
                           ( U(x_i, y_j, z_l) - U_0 )^2}
                                {{\cal N}_{bin}^3 - 1},  
\label{DU}
\end{equation}
where $U_0={\rm M}[U]$ and in our case $U_0={\rm M}[G]=100$.

We introduce the normalized 3D random field 
\begin{equation}
u({\bf r}) = {U({\bf r}) - U_0 \over \sqrt{D[G]}} = 
             \sqrt{U_0}\  \delta ({\bf r}),
\label{u}
\end{equation}

where $\delta ({\bf r}) = [n({\bf r}) - n_0] / n_0$ 
is a standard density contrast,\  $n({\bf r}) = U ({\bf r}) / \Delta_b^3$
is a number density in different cubic bins with coordinates 
designated by a radius-vector ${\bf r}$,\  $n_0 = U_0/ \Delta_b^3$
is a mean number density over the treated volume. 
The denominator is selected  somewhat artificially by analogy with
the Poisson-like  statistic, although we have  $U_0 = {\rm D}[G] \neq {\rm D}[U]$; 
this choice will be justified below.

The value $u({\bf r})$ allows  to calculate a normalized Fourier
transform:
\begin{equation}
F_u ({\bf k}) = {1 \over  L^3}\, \int \, {\rm d}^3{\bf r}\, 
                        u({\bf r})\, e^{-i{\bf k}{\bf r}} = 
              {F_G ({\bf k})\ \sqrt{f_{mod} (k)} \over \sqrt{{\rm D}[G]}},
\label{uk}
\end{equation}
where the second equality follows from Eqs.~(\ref{U}) and (\ref{u})
at $k>0$.
Then  3D power spectrum is:
\begin{eqnarray}
P_{3D}(k) & = & \langle |F_u (k) |^2 \rangle_{V_k} = {1 \over V_k}\ \int_{V_k}\ 
                          {\rm d}^3 {\bf k}'\ |F_u ({\bf k}')|^2 
\nonumber  \\
          & = &  U_0  P_{\rm st} (k),
\label{P3D}
\end{eqnarray}
where   
$\langle...\rangle_{V_k}$ is averaging over a spherical shell in $k$-space,
$V_k=4\pi k^2 \Delta_k$  and $\Delta_k$ 
are the volume and the width of the shell, respectively. 
The integration in Eq.~(\ref{P3D}) 
is equivalent to averaging over   
all possible directions of the vector ${\bf k}$
within the spherical shell $\Delta_k$
(e.g., \citealt{fkp94});
$P_{\rm st}(k) =\,  \langle | \delta_{\bf k} |^2 \rangle$ 
is the standard dimensionless
power spectrum, where  $\langle...\rangle$ is an ensemble averaging  
(e.g., \citealt{p03}).  Using the second equality in Eq.~(\ref{uk})
and Parseval's theorem (e.g., \citealt{fkp94}), applied
to the Gaussian field $G$, 
one can  rewrite Eq.~(\ref{P3D}) as
\begin{equation}
  P_{3D}(k) = {f_{mod}(k) \over {\rm D}[G]}\  
            \langle |F_G ({\bf k})|^2 \rangle_{V_k} \simeq f_{mod}(k).   
\label{P3Df}
\end{equation}
%

It makes sense to introduce also
a normalized correlation (auto-correlation) function 
of the isotropic and 
homogeneous modulated 
random field  $u({\bf r})$:
\begin{eqnarray}
\xi_{3D}(\delta r) & = & \langle u({\bf r}) u({\bf r}\ +\ \delta{\bf r})\rangle_{\rm V}  
\nonumber   \\
                   & = &   {L^3 \over 2 \pi^2}\, \int {\rm d}k\, k\,
                           {\sin{k \delta r}  \over  \delta r}\, P_{3D}(k), 
\label{xi3D}
\end{eqnarray}
where 
$\langle...\rangle_{\rm V}$ -- averaging  
over the normalization  volume  
$V=L^3$, 
typical $\delta r=$
$\sqrt{\delta r_x^2 + \delta r_y^2 + \delta r_z^2}$ 
obeys to a condition  $\delta r << L$.

By analogy with straightforward 3D definitions 
of Eqs.(\ref{u})--(\ref{xi3D})
one can derive also the key quantities for the radial distribution
of objects.  It is convenient to introduce an arbitrary 
centre of coordinates ${\bf r}_0\,  (x_0, y_0, z_0)$,
and concentric spherical layers (bins) $r_b \pm \Delta_r/2$  
at the distance $r_b$ from the centre,  
$r_b=\sqrt{(x_b-x_0)^2+(y_b-y_0)^2+(z_b-z_0)^2}$,
$x_b, y_b, z_b$ -- coordinates 
of points located in  
centres of cubic bins constituting 
a spherical layer,  
$\Delta_r=\Delta_b$ is a width of a layer,
$\Delta_b$ is determined in Eq.~(\ref{FG}).

Let us treat the discrete numbers (in each cell
of the matrix)  
defined by the modulated 
random field $U({\bf r})$,  
as a sample of statistical points (e.g., galaxies).
Then we can calculate 
a radial distribution function, 
$N_R(r)$, as a number of points inside a concentric 
non-overlapping bin: 
\begin{equation}   
N_R(r_b)   =   {1 \over \Delta_b^3}  \int_{r_b - \Delta_r/2}^{r_b + \Delta_r/2} \, 
               {\rm d}r\, r^2\  \int \ {\rm d}\phi\, 
                                  {\rm d}\theta\, 
                                  \sin{\theta}\, U({\bf r}),           
\label{NR}
\end{equation}
where 
$n(r_b) = N_R(r_b)/4\pi r_b^2 \Delta_r$
$\approx n(r)$ 
is a mean number density
within a concentric layer.

In analogy with Eq.~(\ref{u})
we introduce a normalized 
radial distribution function:
\begin{equation}
u_R(r_b) = {N_R(r_b) \, - \,  n_0\ 4\pi\ r_b^2\, 
                           \Delta_r  \over 
                            \sqrt{ n_0\ 4\pi\ r_b^2\, \Delta_r}}
                     = \langle u({\bf r}) \rangle_{r_b} \sqrt{{\cal N}_{r_b}},   
\label{uR}
\end{equation}
where  $n_0\ =\ U_0/{\Delta_b^3}$,\ 
$\langle u({\bf r}) \rangle_{r_b}$ is the normalized random
field (\ref{u}) averaged over the spherical layer;   
${\cal N}_{r_b}$ 
is a number of cubic bins inside the spherical shell;
at ${\cal N}_{r_b} \gg 1$ we have 
${\cal N}_{r_b} = 4\pi r_b^2 \Delta_r/ \Delta_b^3$. 
The difference between Eqs.~(\ref{u})
and (\ref{uR}) is that  (\ref{u}) refers to
any points ${\bf r}$ in the considered volume while 
(\ref{uR}) describes  spherical layers around a
selected centre.

The radial distributions of real objects (e.g., 
luminous red galaxies (LRG) discussed in Paper~I) 
display quite complex
radial behaviours including large-scale variations or
so-called {\it trends}.  
To study specially  intermediate  scales
one needs to reduce an influence of the
largest scales or the smallest wave numbers $k$ 
in the Fourier space.
Therefore, in some cases we 
introduce a trend-subtraction procedure
which consists in a replacement of the value
$n_0 4\pi r_b^2 \Delta_r$ in Eq.~(\ref{uR})   
by a smooth 
{\it trend} function  $N_{\rm tr}(r_b)$.
Such  modified forms of Eq.~(\ref{uR}) are specially 
indicated  in the text. 

In principle 
the normalized radial distribution $u_R(r_b)$ 
can be treated as a 3D-distribution averaged over
angles within each shell  and 
with the same variance as $u({\bf r})$ in (\ref{u}).
Actually, it can be shown that ${\rm D}[u_R]={\rm D}[u]$.
However,
here we focus specially on 1D-approach to the radial
distribution of $u_R (r_b)$ as more simple  way to
detect possible variations in spatial distributions 
of the cosmologically remote objects.
Thus we  calculate the power spectrum
of the normalized radial distribution 
$u_R(r_b)$ using the one-dimensional definition
(e.g., \citealt{jw69,  sc82})
\begin{eqnarray}
\label{PRk}
& &   P_R (k_m) = |F_R^{1D} (k_m)|^2 = 
\nonumber                   \\
& &
{1 \over {\cal N}_{R, bin}}   
                     \left\{ \left[
          \sum_{j=1}^{{\cal N}_{R, bin}}  u_{R, j}  
	  \cos(k_m r_{b, j})
              \right]^2   +
              \right.     
\nonumber   \\ 
& &  \left.  
\left[ \sum_{j=1}^{{\cal N}_{R, bin}}  
	       u_{R, j}
             \sin (k_m r_{b, j})
             \right]^2 \right\},
\end{eqnarray}
where 
$ F_R^{1D} (k_m) = ({\cal N}_{R, bin})^{-1/2}   
\sum_{j=1}^{{\cal N}_{R, bin}}  u_{R, j} e^{-i k_m r_{b, j}}$
is the one-dimensional discrete  Fourier transform,
${\cal N}_{R, bin}$ is a number of concentric bins (shells), 
$j=1,2,...\ {\cal N}_{R, bin}$ is a numeration of the  bins, 
$r_{b, j}$ is a localization of a centre of $j$-th  bin,
$k_m=2\pi  m/L_R$ is a wave number
corresponding to an integer harmonic number
$m=1,2,...{\cal M}$,\  ${\cal M}={\cal N}_{R, bin}/2$ is
a maximal number (the Nyquist number) 
of independent discrete harmonics,
$L_R$  is the whole interval 
in the configuration space, i.e., so-called sampling length. 

The value $P_R (k)$ is the key  value
of our radial approach, where $k$ is conjugate value
to the radial distance $r$.
Note that the power spectrum (\ref{PRk}) 
differs from the power spectrum  $P_{1D}(k_z)$  of the pencil-beam 
1D samples  discussed, e.g. by \cite{kp91},
because as opposed to a single direction 
in the case of $P_{1D}(k_z)$
in (\ref{NR}),\ (\ref{uR}) 
we analyze a variety of directions 
(averaged over the angles)
characterized by
the single variable $r_b$  measured  relative 
to the selected  centre. 

Following an analogy with Eq.~(\ref{xi3D}), but 
employing the one-dimensional approach of Eq.(\ref{PRk}), and 
continuing the discrete consideration of Eqs.~(\ref{NR}), (\ref{uR}) 
and (\ref{PRk}) we can introduce a value:

\begin{eqnarray}
\label{xiR}
 \xi_R(\delta r) & = & \langle u_R(r_b)\ u_R(r_b + \delta r) \rangle_{\rm R} 
\nonumber                                                        \\
     & = &  2 \sum_{m=1}^{\cal M}\  P_R (k_m)\,  \cos ( k_m \delta r),
\end{eqnarray}
where $\delta r$ designates a set of discrete distances between 
central radii $r_{b, j}$ of concentric layers,\, 
$\langle ... \rangle_{\rm R}= \sum_{j=1}^{{\cal N}_{R, bin}} .../ {\cal N}_{R, bin}$ 
is an averaging  over 
all concentric layers 
consistent with the sampling length $L_R$;
this is similar  to the averaging $\langle ... \rangle_{\rm V}$
introduced in Eq.~(\ref{xi3D}) but with central 
symmetry preservation.

\section{MODEL SIMULATIONS}
\label{sec:ms}  
The main goal of this work is to carry out simulations 
of the modulated  Gaussian 3D-fields
having a chance to reproduce -- with some probability --
quasi-oscillations in the radial 
(shell-like) distributions of 
cosmological objects. 

To trace  the modulation effects 
on properties 
of the radial distribution of matter 
we introduce a
modulating function
$f_{\rm mod}(k) = f_1 (k) + f_2 (k)$, 
where a smooth function $f_1 (k)$ is designed as 
\begin{equation}
f_1 (k) = f_{\rm CDM} (k) + 1,
\label{f1}
\end{equation}
here  $f_{\rm CDM}(k)$   
is a power spectrum of the cold dark matter (CDM) density 
(e.g., \citealt{bbks86}) 
\begin{equation}
f_{\rm CDM} (k) = A_1 \cdot q\,  {\rm T}^2(q),
\label{fCDM}
\end{equation}
$A_1$ is a normalizing constant,
$q$ is dimensionless 
variable $k=|{\bf k}|$
determined according to \citet{s95} as 
\begin{equation}
q=\frac{k/({\rm Mpc}^{-1}\ h)}{\Omega_{\rm m}  h  
\exp[-\Omega_{\rm b}(1 + \sqrt{2 h}/\Omega_{\rm m})]},
\label{q}
\end{equation}
where $\Omega_{\rm m}$ is the relative total density of matter,\ 
$\Omega_{\rm b}$ is the relative density of baryons,\
${\rm T}(q)$ is a transfer function: 
\begin{eqnarray}
\label{Tq}
& & {\rm T}(q)={\ln (1+2.34 q) \over 2.34 q} \times  \\
\nonumber    
& & [1+3.89 q + (16.1 q)^2 + 
                         (5.46 q)^3 + (6.71 q)^4]^{-1/4}.
\end{eqnarray}
The second term 
 ``1''  in the right side of  Eq.~(\ref{f1})
stands for so called ``shot noise'' (e.g., \citealt{fkp94}),
which dominates at small scales (large $k$).

The function $f_2(k)$ is simulated as a product 
\begin{equation}
f_2(k) = f_{\rm CDM}(k)\, \cdot  f_{\rm BAO}(k), 
\label{f2}
\end{equation}
where $f_{\rm CDM} (k)$ is defined by Eqs.~(\ref{fCDM})-(\ref{Tq})
and  $f_{\rm BAO} (k)$ 
is a  damped oscillation function
designed as a modification of the fitting formula (3) 
of \citet{bg03} introduced to imitate the BAOs:
\begin{eqnarray}
& & f_{\rm BAO}(k) = A_2\  {k \over k_s} \  \exp 
          \left[- \left({k \over 0.1~h~{\rm Mpc}^{-1}}\right)^{1.4} \right] \times 
\nonumber                       \\
& &       \left[1+ \sin\ ({2\pi k \over k_s} + \phi_s)\right], 
\label{fBAO}
\end{eqnarray}
where $k_s=0.0628\ h$~Mpc$^{-1}$ or $2\pi/k_s=100~h^{-1}$~Mpc, 
$\phi_s$ is a phase; below we choose 
$\phi_s=1.4$ to 
make a position of the main peak of $f_2(k)$ 
equal to $k_s$. In what follows one can make sure that the 
accurate form of Eq.~(\ref{fBAO}) is not important.

\begin{figure*}[htb!]   
\begin{center}
\includegraphics[width=0.95\textwidth]{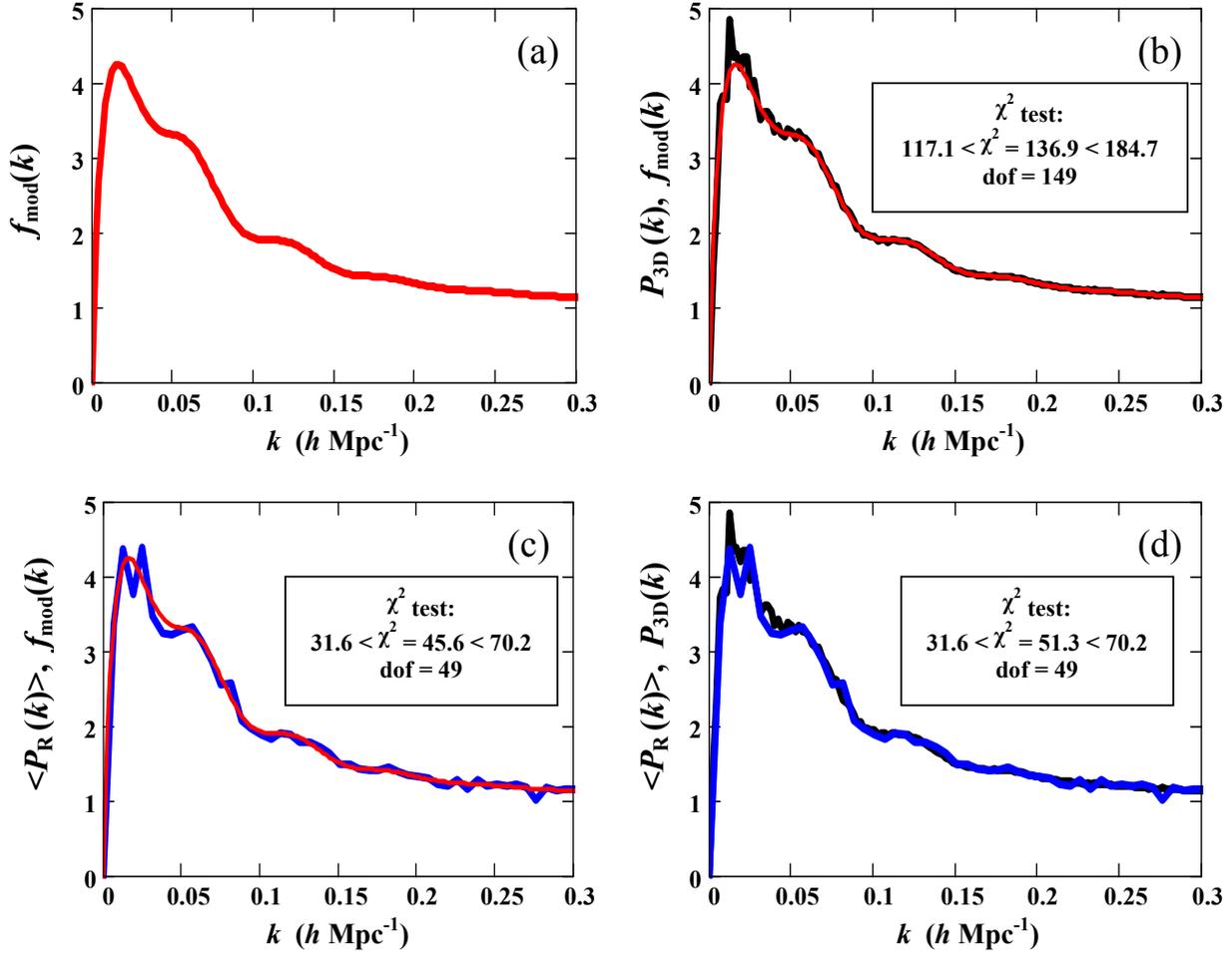}
\caption{
(Colour online)
{\it Panel (a)} $-$ function $f_{\rm mod}=f_1 (k) + f_2 (k)$   ({\it solid})
used for modulation of the 3D Gaussian 
field in  ${\bf k}$-space, where
$f_1$ is a  model function simulating
a smoothed 3D power spectrum 
[Eq.~\protect{(\ref{f1})}],
$f_2$ is a damped oscillating model function 
[Eq.~\protect{(\ref{f2})}];\,  
{\it panel (b)} $-$ comparison of simulated 
power spectrum of the 3D Gaussian 
field $P_{3D}(k)$ ({\it thick wavy curve}) 
[see text] with modulating 
function $f_{\rm mod}(k)$ ({\it thinner solid curve}) ; 
the insert shows result
of $\chi^2$ two-sides test calculations for 150 bins
along $k$-axis.  A simulated quantity
of the Pearson sample statistics  
$\chi^2 = \chi_{\nu}^2$, where 
$\nu = 149$ is a number of degrees-of-freedom (dof), and 
critical quantiles confining a confidence region
(with confidence probability $0.95$)  
are represented;\,
{\it panel (c)} $-$ comparison of simulated 
radial power spectrum $\langle P_{R}(k) \rangle$  
({\it thick solid curve}) averaged over 500 radial power spectra 
with modulating  function $f_{\rm mod}(k)$ ({\it thinner solid curve});
{\it panel (d)} $-$ comparison of $P_{3D}(k)$ (same as in panel (b))
with $\langle P_{R}(k) \rangle$  (same as in panel (c));\, 
inserts in the panels (c) and (d) are organized similar 
to the insert in the panel (b) but for simulations
with $\nu = 49$ dof.  
}  
\label{f1f2}
\end{center}
\end{figure*}
The modulating function
\begin{equation}
f_{\rm mod}(k)=f_{\rm CDM}(k) \cdot [1+f_{\rm BAO}] + 1 
\label{fmod}
\end{equation}
is represented in Fig.~\ref{f1f2}(a). 
For certainty we put the normalizing  
constant $A_1=470$ as a better approximation
to the observational power spectrum 
(used from \citealt{anderetal14}). 
Note that the functions 
$f_1 (k) - 1 = f_{\rm CDM}(k)$ and $f_2(k)$ tend
to zero at $k \rightarrow 0$, 
i.e at very large scales.

Additionally, we introduce 
$A_m(k)=f_2(k)/f_1(k)$ as a relative  amplitude of 
oscillating modulation in dependence on $k$.
In particular, the main peak of the modulating function
$f_2(k_s)$  
corresponds to 
$A_m = A_m (k_s)=f_{\rm BAO}(k_s) \cdot 
f_{\rm CDM}(k_s)\ [f_{\rm CDM}(k_s)+1]^{-1}$.
To make the effects of the BAO more visible
we admit hereafter 
that $A_m=0.2$. 
It is approximately twice as much as
the double observational BAO amplitude
($\approx  2 \times 0.05 = 0.1$, e.g.,
\citealt{bg03},\  \citealt{anderetal14}). 
This way we overestimate effects 
of the  oscillating part of the modulation 
on the radial power spectra. 

\begin{figure*}[htb!]   
\begin{center}
\includegraphics[width=0.95\textwidth]{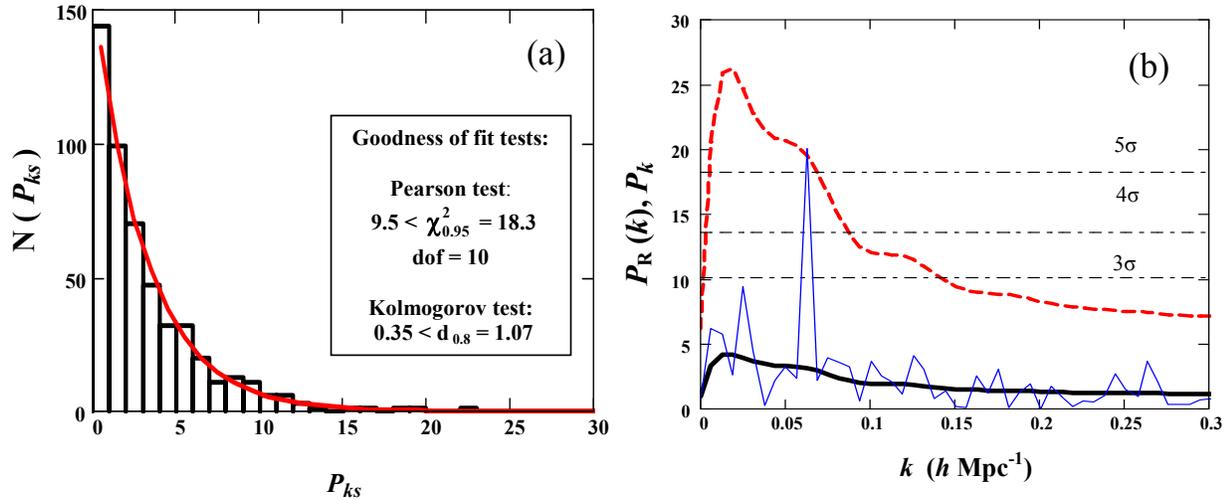}
\caption{
{\it Panel (a)} $-$ histogram of 
peak values $P_{ks}$ at fixed 
$k=k_s=0.0628~h$~Mpc$^{-1}$\,  obtained for the
same 500 radial distributions as in 
Fig.~\ref{f1f2} (c) and (d), 
within  bins  $n \leq P_{ks} \leq n+1$, 
where $n=0, 1, 2,..., n_{\rm max}$, and compiled with
an exponential distribution 
of peaks ({\it solid line}).   
Insert shows results of  Pearson's $\chi$-square 
and Kolmogorov's goodness of fit tests.
{\it Panel (b)} $-$ power spectrum 
$P_R (k)$ ({\it thin  broken} curve) 
calculated for
one chosen realization of radial distributions 
with  rather prominent 
peak near $k_s$;\   {\it thick solid} line --
the same modulation function $f_{\rm mod} (k)$
as in Fig.~\ref{f1f2}~(a),
but shown in  different vertical scale;
{\it thick dashed} line is the confidence level
$\beta=0.998$  drawn  using 
the mean amplitudes $f_{\rm mod} (k)$ and
the exponential distribution function  
\protect{(\ref{calF})};
the horizontal {\it thin dot-dashed} lines 
show significance levels 
estimated according to \cite{sc82}  
(see text for details).
}
\label{3sigma}
\end{center}
\end{figure*}

\begin{figure*}[htb!]   
\begin{center}
\includegraphics[width=0.95\textwidth]{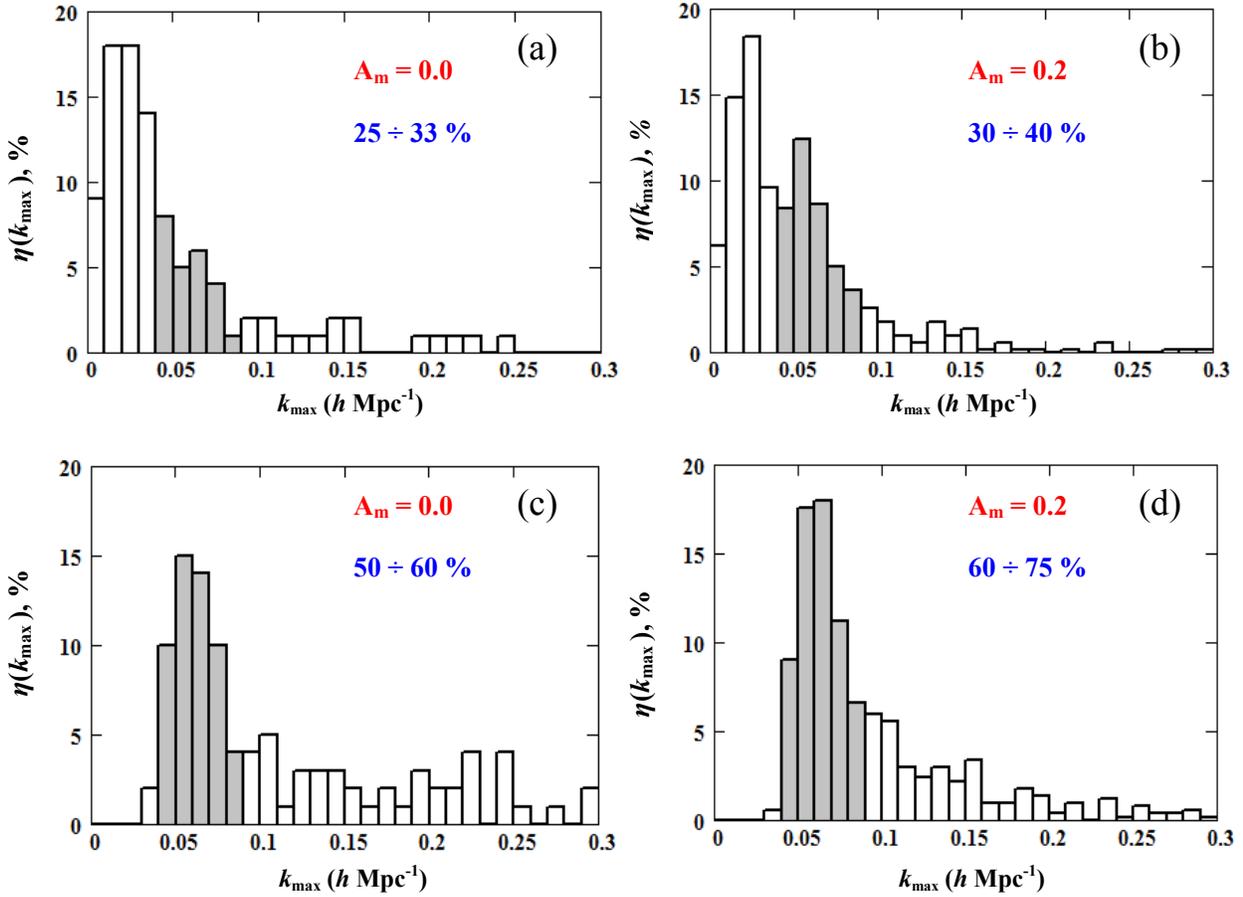}
\caption{
Histograms displaying
a frequency of occurrence
$\eta(k_{\rm max})$\   \protect{(\ref{eta})} 
to find the main peak positions $k_{\rm max}$
of radial power spectra 
within  independent   
bins $\Delta k_{\rm max} = 0.01~h$~Mpc$^{-1}$;
dark grey columns indicate regions 
$0.04 \leq k_{\rm max} \leq 0.09~h$~Mpc$^{-1}$. 
All four panels are plotted for 
a sample length  $L_R = 1000~h^{-1}$~Mpc;\,
{\it  panels~(a)} and {\it (c)} 
correspond to the case
of  zero amplitude  $A_m=0$ (no BAO effects);
{\it panels~(b)} and {\it (d)} -- to $A_m=0.2$.
Approximate fractions (in percents) 
of peak positions $k_{\rm max}$
falling into the dark grey columns  
are indicated in all panels.
{\it Panels~(c)} and {\it (d)} display effects of 
a trend subtraction (see text for details)
extinguishing Fourier components with  $k < 0.04~h$~Mpc$^{-1}$.
}
\label{Histo}
\end{center}
\end{figure*}

\begin{figure*}[htb!]   
\begin{center}
\includegraphics[width=0.95\textwidth]{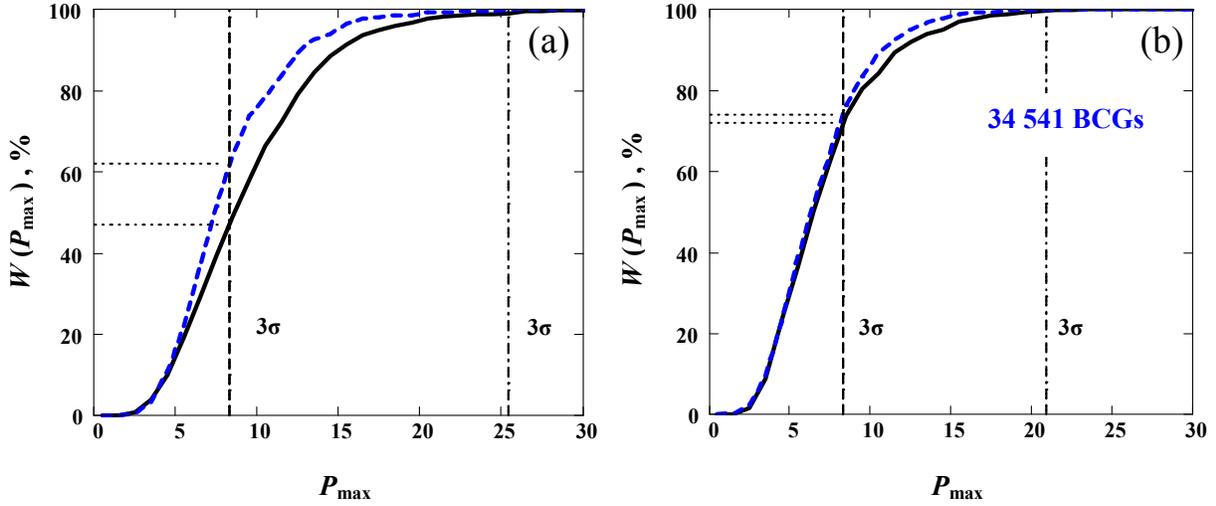}
\caption{
(Colour online)
Cumulative probability 
$W(P_{\rm max}) = W(P_{k} \leq P_{\rm max})$ 
to reveal a dominant peak of  power spectrum   
within the interval 
$0.04 \leq k \leq 0.09~h$~Mpc$^{-1}$ 
with an amplitude $P_k$ 
not exceeding a fixed value $P_{\rm max}$. 
The vertical {\it shot dashed} lines 
indicate significance levels $3 \sigma$ 
(slightly  higher)  
corresponding to values $P_{\rm max}$ 
estimated according to  \citet{sc82};\
the vertical {\it dot-dashed} lines show
the same significance levels 
but determined 
by a condition $W(P_{\rm max}) = 99.8\%$
(see text);\
the horizontal short  dashed lines show
the levels above which 
significance of the peak exceeds $3\sigma$ 
(following to \citealt{sc82})  
with probability  $1-W(P_{\rm max})$. 
All curves $W(P_{\rm max})$  
refer to the sampling length -- $1000~h^{-1}$~Mpc.
{\it Panel (a)} $-$ 
{\it solid} curve corresponds 
to the amplitude $A_m=0.2$, 
{\it dashed} curve $-$ to  $A_m=0$.
{\it Panel (b)} $-$  {\it solid} curve 
represents $W(P_{\rm max})$ calculated for
34~541 brightest cluster galaxies (BCGs)
(see Sect~\ref{sec:bcg}),  {\it dashed}
curve is simulated  at $A_m=0$ 
similar to the {\it dashed}
curve in the panel (a),
but produced
for a reduced constant  
$A_1' =  340$ 
in \protect{Eq.~(\ref{fCDM})}.
}  
\label{WPmax}
\end{center}
\end{figure*}

Employing Eqs.~(\ref{U}), (\ref{NR}) and (\ref{uR}) 
we produce the modulated random fields in the Cartesian coordinate
system (CS) and  calculate 
a set of normalized  
radial (shell-like) distribution functions $u_{\rm R}(r)$,
where $r$ 
is a distance between 
coordinates  ($x, y, z$) 
of field  points   
and arbitrary chosen centres (null points) 
of the radial distributions  
${\bf r}_0$\  ($x_0, y_0, z_0$).
Using Eq.~(\ref{PRk}) we obtain 
a sample of radial  power spectra $P_{\rm R}(k)$,
where $k$ is a variable conjugate to the distance $r$,
and carry out a statistical analysis of the sample.  
The results of such analysis are represented in 
Figs.~\ref{f1f2}--\ref{WPmax}.

So the thick solid curve in 
Fig.~\ref{f1f2}(a) displays the modulating
function $f_{\rm mod}=f_1 (k) + f_2 (k)$  
introduced in Eq.~(\ref{fmod})
to simulate the  modulation of the 3D Gaussian field
in  ${\bf k}$-space.  
A comparison of simulated 3D
power spectrum  $P_{3D}(k)$
obtained with using Eq.~(\ref{P3D}) 
and the modulating 
function $f_{\rm mod}(k)$ is represented in Fig.~\ref{f1f2}(b). 
The insert shows results
of $\chi^2$ two-sides test calculations for 150 bins
along $k$-axis. In this case we use a definition
\begin{eqnarray}
\chi^2=\chi_{\nu}^2 & = & \sum_{m=1}^{\cal M} 
        \left( { P_{3D} (k_m) - f_{\rm mod} (k_m) 
        \over \sigma_{3D}(k_m) } \right)^2, 
\nonumber  \\
         \sigma_{3D} (k_m) & = &  P_{3D} (k_m)\ N_m^{-1/2},
\label{fig1b}
\end{eqnarray}
where ${\cal M}$ is the Nyquist number (${\cal M} = 150$),\quad  
$\sigma_{3D} (k_m)$ is the standard deviation 
determined for 
statistics of the mean  values,  
$N_m$ is the number of values
$P_{3D} (k_m)$ involved in averaging 
within a spherical shell (\ref{P3D})
$V_k=4\pi k_m^2 \Delta_k$, i.e., approximately 
$N_m \approx 4\pi (k_m/\Delta_k)^2$.
Note that
the sample length is chosen as $L=3000$~Mpc~$h^{-1}$ and 
$\Delta_k=2\pi/L \simeq 0.0021~h$~Mpc$^{-1}$.

A comparison of simulated radial
power spectrum $\langle P_R (k) \rangle$  
averaged over 500 power spectra 
calculated for different radial distributions   
with the modulating  function 
$f_{\rm mod}(k)$ is shown in Fig.~\ref{f1f2} (c).
The insert also shows the results
of $\chi^2$ two-sides test calculations but for 50 bins
along $k$-axis, in this case we use 
\begin{eqnarray}
\chi^2=\chi_{\nu}^2 & = & \sum_{m=1}^{\cal M} 
        \left( {\langle P_R (k_m) \rangle - f_{\rm mod}(k_m) 
         \over \sigma_R (k_m) } \right)^2,
\nonumber    \\
         \sigma_R (k_m) & = & \langle P_R (k_m) \rangle\ N_{rd}^{-1/2},
\label{fig1c}
\end{eqnarray}
where ${\cal M}=50$,\quad
$\sigma_R (k_m)$ is the standard deviation
determined similar to Eq.~(\ref{fig1b}),  
$ N_{rd}=500 $ is the number of radial distributions, here
the sample length is $L_R=1000$~Mpc~$h^{-1}$ and 
$\Delta_k=2\pi/L_R  \simeq 0.0063~h$~Mpc$^{-1}$.

Fig.~\ref{f1f2} (d) represents a comparison  
of $P_{3D}(k)$ [same as in the panel (b)]
with $\langle P_R(k) \rangle$  [same as in the panel (c)]. 
The results 
of $\chi^2$ two-sides test calculations  for 50 bins
along $k$-axis are shown in the insert. Carrying out 
these calculations we use
\begin{eqnarray}
\chi^2=\chi_{\nu}^2 & = & \sum_{m=1}^{\cal M} 
        \left( { \langle P_R (k_m) \rangle - P_{3D} (k_m) 
         \over \sigma_{sm} (k_m) } \right)^2, 
\nonumber     \\
         \sigma_{sm}(k_m) & = & \sqrt{\sigma_{3D} (k_m)^2 + \sigma_R(k_m)^2},
\label{fig1d}
\end{eqnarray}
where ${\cal M}=50$,\quad  $\sigma_{sm} (k_m)$ is the total standard deviation
determined by  Eqs.~(\ref{fig1b}) and (\ref{fig1c}).  
Here as in Fig.~\ref{f1f2} (c)
the sampling length is $L_R = 1000$~Mpc~$h^{-1}$ for simulations of
$\langle P_R (k) \rangle$ 
and as in Fig.~\ref{f1f2} (b) -- 
$L=3000$~Mpc~$h^{-1}$ for simulations of $P_{3D} (k)$.
In such a way we use  the same bin in $k$-space   
$\Delta_k \simeq 0.0063~h$~Mpc$^{-1}$ 
for calculations of $\langle P_R (k) \rangle$
and every third bin $\Delta_k  \simeq 0.0021~h$~Mpc$^{-1}$ 
for calculations of $P_{3D} (k)$.

One can see that
the results of calculations represented in Fig.~\ref{f1f2}
do not contradict (with confident probability 0.95)
the hypothesis which can be
formalized as an asymptotic double equality
\begin{equation}
f_{\rm mod}(k) = P_{3D}(k) =  \langle  P_R (k) \rangle,
\label{maineq}
\end{equation}
where $\langle ... \rangle = \langle...\rangle_{ens}$ 
is an ensemble averaging, i.e.
the averaging over a set of 
respective radial distributions
with numerous centres in real space.

The first equality in Eq.~(\ref{maineq}) 
is consistent
with Eq.~(\ref{P3Df}),  the second one 
can be roughly explained as follows.
If we produce additional averaging of 
the value  $|F_R^{1D} (k)|^2$ given in Eq.~(\ref{PRk})
over an interval $k-\Delta_k/2 \leq k \leq k+\Delta_k$
in each realization of the normalized random field $u_R (r)$
(considering $k$ as a continuous variable)
and implement an ensemble averaging in the $k$-space, 
then we can assume that 
$\langle \langle |F_R^{1D} (k)|^2 \rangle_{\Delta_k} \rangle_{ens} =
\langle |F_R (k)|^2 \rangle_{V_k}$,
where $\langle ...\rangle_{V_k}$
is determined in Eq.~(\ref{P3D}) and
$F_R(k)$ is the 3D Fourier transform of $u_R (r)$. 
In such a case we can expect
that $\langle |F_R (k)|^2 \rangle_{V_k} = \langle |F_u (k)|^2 \rangle_{V_k}$,
because both the fields $u_R$ and $u$ 
have zero mean values and the same variance;  
the last  equality brings us back to the Eq.~(\ref{P3Df}).
These assumptions have been confirmed by our simulations.%
\footnote{Note that another denominator in Eq.~(\ref{u}),
e.g. $({\rm D}[U])^{1/2}$,
would lead to an additional factor in the last equality 
of Eq.~(\ref{P3Df}),
e.g. $(U_0/{\rm D}[U])$,
which could also be taken into 
account in our analysis.}  

The second equality in Eq.~(\ref{maineq})
leads to an unexpected consequence, which 
can be verified. Actually, if we produce ensemble  
averaging $\langle ... \rangle_{ens}$ of Eq.~(\ref{xiR}), then
with using  Eq.~(\ref{maineq}) we get
(in the approximation of continuous variables)
\begin{equation}
\langle \xi_R(\delta r) \rangle = {L \over \pi} \int_0^\infty {\rm d} k \cos(k \delta r) P_{3D}(k).
\label{avxiR}
\end{equation}
Comparing Eqs.~(\ref{avxiR}) and (\ref{xi3D}) one can finally obtain
\begin{equation}
\langle \xi_R(\delta r) \rangle = {2\pi \over L^2}\ \int_{\delta r}^\infty
                             {\rm d} y \  y\ \xi_{3D}(y),
\label{xiRxi3D}
\end{equation}
where $\delta r$ is a distance along an arbitrary radial direction.%
\footnote{The right hand side of  
Eq.~(\ref{xiRxi3D}) can be interpreted 
as an integral over 
all directions $\delta { \bf r}_\perp$ 
transverse to an  arbitrary  radial direction 
$\delta r = \delta r_\parallel$
so that $y = \sqrt{(\delta r_\perp)^2 + (\delta r_\parallel)^2}$.  
The equality (\ref{xiRxi3D}) is to some extent 
complementary  to  Eq.~(4.8) by \cite{kp91} which
refers  to the relationship between the
one-dimensional power spectrum $P_{1D}(k_\parallel)$ 
along a beam direction $k_\parallel$
and the three-dimensional power spectrum 
$P_{3D}(k)$, where $k=\sqrt{(k_\perp)^2 + (k_\parallel)^2}$. 
In the latter case the equality 
$\xi_{3D}(\delta r) = \xi_{1D}(\delta r)$
is valid for any chosen directions.} 

Our numerous simulations of radial distributions
and their power spectra $P_R (k)$   have shown that
peak height (amplitudes) $P_{k}$
at any fixed  $k$  are distributed
according to an exponential distribution
with the mean (mathematical expectation)
peak amplitudes  M$[P_{k}]$.
In its turn the  mean value M$[P_{k}]$  
can be determined as the
peak amplitude $\langle P_R (k) \rangle$
averaged over many centres of radial distributions
or,  according to Eq.~(\ref{maineq}), 
as $f_{\rm mod}(k)$.

Consequently 
the cumulative  distribution function 
integrating over all values of
peak amplitudes $P_{k}$  lower than a fixed
value  $P^*_{k}$  can be expressed as
\begin{equation}
{\cal F}(P_{k} < P^*_{k},\  \lambda) = 1 - \exp(-\lambda\ \cdot P^*_{k} )
\, \, \, \, {\rm at} \, \, \, \,  P^*_{k}  \geq  0,
\label{calF}
\end{equation}
where $\lambda = \lambda(k)$ 
is a parameter of the exponential distribution
determined by a reciprocal 
modulating function,
i.e.  $\lambda (k) = f_{\rm mod}^{-1} (k)$. 
Let us emphasize that the difference between 
Eq.~(13) of \citet{sc82} or Eq.~(7) of \citet{fef08} 
and  equation (\ref{calF}) is a  constant
parameter $\lambda$ of the exponential distributions against 
a variable $\lambda (k)$ in the present study.

Fig.~\ref{3sigma}(a) demonstrates 
a histogram  of numbers N$(P_{ks})$ 
of peak values  
falling in a bin  $n \leq P_{ks} \leq n+1$, \
$n=0, 1, 2, ..., n_{\rm max}$, 
at fixed  $k=k_s=0.0628~h$~Mpc$^{-1}$ 
obtained for the same set of
radial distributions as in Fig.~\ref{f1f2}.  
The solid line 
shows theoretical function 
N$_{th}(P_{ks})$ of 
peak amplitudes within 
each bin at
$\lambda (k=k_s) = 1/3.14 = 0.32$  
(where $f_{\rm mod} (k_s)=3.14$ at $A_1=470$)
calculated with the use of 
Eq.~(\ref{calF})
following a formula 
\begin{equation}
{\rm N}_{th}(P_{ks}) = [{\cal F} (P_{ks} < n+1) - {\cal F} (P_{ks} < n)] \cdot 500, 
\label{calFcalF}
\end{equation}
where ``500'' stands for the number of trials.

The insert gives results of two one-side 
goodness-of-fit tests  
between the smooth
theoretical curve (\ref{calFcalF})
and the histogram: 
Pearson's $\chi$-square and Kolmogorov's tests
(e.g., \citealt{ll84}). 
Both the tests  demonstrate good agreements 
at the confidence level 0.95 for $\chi$-square test
and 0.8 for the Kolmogorov's test, the sample size
in the former case is $n_{\rm max}=13$ and dof$=13-3=10$
(the number 3 corresponds to the most stringent criterion), 
while in the latter one the sample  volume is $n_{\rm max}=23$; 
respective 
quantiles  
are also indicated in the insert.

Fig.~\ref{3sigma}(b)
demonstrates an example of $P_R (k)$  (thin solid curve)
calculated  using
Eq.~(\ref{PRk}) 
for one realization 
of the modulated random field $U({\bf r})$\  [Eq.~(\ref{U})]
and,  accordingly, of $u_R(r)$\  [Eq.~(\ref{uR})]
at the sampling length
$L_R=1000~h^{-1}$~Mpc  (${\cal N}_{R, bin}=100$). 
One can see a noticeable peak in the power 
spectra at $k=k_{\rm max}$ 
which is approximately equal to
$k_s=0.0628~h$~Mpc$^{-1}$. 
The peak corresponds 
to a scale of quasi-periodicity
$  \Delta D_c  =  2\pi / k_{\rm max} =  L_R/m_{\rm max} 
\simeq 100~h^{-1}$~Mpc
with $m=m_{\rm max} \simeq 10$.

Employing  Eq.~(\ref{calF}) 
we  produce   estimations of the 
peak amplitudes $P_k$ at a given confidence level $\beta$.
The panel (b) demonstrates such a level at $\beta=0.998$
(slightly higher then  the Gaussian significance $3\sigma$)
for different $k$ (thick dashed curve) together with the 
mean function $f_{\rm mod} (k)$ (thick solid curve). 
The horizontal (thin dot-dashed) lines in the panel (b)
show the conventional significance 
levels   $3\sigma$ , $4\sigma$, and  $5\sigma$
(equivalent to the Gaussian confidence probabilities  
$\beta=0.998,\ 0.999 936,\ 0.999 9994$, respectively)
calculated for the peak amplitudes $P_k$
with the use of the false alarm probability
(\citealt{sc82},\  \citealt{fef08}).
Note that an amplitude of the dominant 
peak in Fig.~\ref{3sigma}(b)
exceeds the level $5\sigma$ according to \citet{sc82} estimations,
but this amplitude turns out to be 
slightly higher 
than the level $3\sigma$ estimated using  the
exponential distribution Eq.~(\ref{calF}). 
In this paper we refer to 
this more stringent  way to estimate the significance of
the relatively high peaks ($P_k \ga 20$) 
in the power spectra which may
occur in the modulated Gaussian field.

Fig.~\ref{Histo} demonstrates a
frequency of occurrence  $\eta(k_{\rm max})$ (in $\%$) 
of  dominate-peak positions $k_{\rm max}$  
within
different independent bins. 
This value  can be  defined as 
\begin{equation}
\eta(k_{\rm max}) = {{\cal N}_{peak}(k_{\rm max}) \over {\cal N}_{\rm real}}, 
\label{eta}
\end{equation}
where ${\cal N}_{peak}(k_{\rm max})$ is a number of highest peaks
falling (independently of their amplitudes)  
within an interval (bin)
$k_{\rm max}^c -\ \Delta k_{\rm max}/2 \leq k_{\rm max} \leq k_{\rm max}^c$ 
$+\ \Delta k_{\rm max}/2$,\ $k_{\rm max}^c$ is a centre of a
bin,\  $\Delta k_{\rm max}=0.01$;\  ${\cal N}_{\rm real}$ 
is a full number of realizations 
(in our case ${\cal N}_{\rm real}=500$);
the sampling length is $L_R=1000~h^{-1}$~Mpc.

Panels (a) and (c) of Fig.~\ref{Histo} 
correspond to zero value of 
modulating function
$A_m=f_2(k_s)/f_1(k_s)$  $=0$
which means zero oscillating component of the modulating 
function $f_{\rm mod} (k) = f_1 (k)$, while 
panels (b) and (d) correspond to $A_m=0.2$.
Panels (c) and (d) of Fig.~\ref{Histo} 
display  the effects of 
the trend-subtraction procedure 
with substitute  the value  
$n_0  4\pi r_b^2 \Delta_r$
in Eq.~(\ref{uR}) 
by  a smoothed function (trend) $N_{\rm tr}(r_b)$ 
filtering out 
the largest scales (Fourier components with 
$k < 0.04~h$~Mpc$^{-1}$).
To plot both the histograms 
in the panels (c) and (d)
we calculate the normalized values $u_R (r_b)$
for all partial radial distributions
with making use of this procedure. 
To calculate 
trend functions $N_{\rm tr}(r_b)$
we employ the least-squares method with using 
a set of parabolas.
Let us emphasize, however, that the
procedure of the trend subtraction 
is quite ambiguous (e.g., Paper~I) and we
use it in our simulations 
only for qualitative consideration.

Dark grey columns in Fig.~\ref{Histo}   
display fractions of 
peaks  falling into an interval 
$0.04 \leq k_{\rm max} \leq 0.09~h$~Mpc$^{-1}$
which corresponds to the main bump
of the modulation function $f_2(k)$
in Fig.~\ref{f1f2}. 
Comparing 
the panels (b) and (a) one can notice a visible 
excess ($\sim 5-10 \%$)
of the occurrence $\eta (k_{\rm max})$  at  $A_m=0.2$
relatively the same value at $A_m=0$.
Remind that in  Fig.~\ref{Histo}  
we ignore the significance 
of highest peaks in the radial power spectra 
being interested only in their positions.
Similar moderate gain ($\sim 10-15\%$) 
in the occurrence of the peaks 
is noticeable 
also from a  comparison of   
the panels (d) and (c).
On the other hand, an artificial suppression of 
the smallest $k$ by the trend elimination  
leads to increase of $\eta(k_{\rm max})$
in the  bins  belonging to
the interval of our interest 
(cf. panels (c) and (a)
at $A_m=0$, as well as  
(d) and (b) at $A_m=0.2$). 
One can notice some  pumping  of the Fourier components
in the nearest regions of  $k \ga 0.04~h$~Mpc$^{-1}$.
In any case, there is  
enhanced probability to find 
quasi-periodicities 
in the radial distributions 
of matter with a scale $2\pi/k$, where $k$
belongs to the considered interval.

Fig.~\ref{WPmax} represents  
a cumulative probability 
(in $\%$) 
$W(P_{\rm max})$ 
to reveal a dominant peak 
in the radial power spectrum
with an amplitude $P_k$
not exceeding  a fixed value $P_{\rm max}$
at $k=k_{\rm max}$
falling into
the same range $0.04 \leq k_{\rm max} \leq 0.09~h$~Mpc$^{-1}$.
This value differs from Eq.~(\ref{calF})  by relatively
wide range of values $k$ instead of fixed $k$ used
in (\ref{calF}). Both the panels in
Fig.~\ref{WPmax} is plotted for  samples 
of 500 radial power spectra at either  $A_m=0.0$ or $0.2$ 
calculated for $k$  from
the whole range of 
$0.0 \leq k \leq 0.3~h$~Mpc$^{-1}$.
We evaluate   $W(P_{\rm max})$   
employing our model simulations or observational data
(see below) using natural definition
\begin{equation}
W(P_{\rm max})={{\cal N}(P_k \leq P_{\rm max}) \over {\cal N}_{\rm real}},
\label{WP}
\end{equation}
where ${\cal N}(P_k \leq P_{\rm max})$ is a full number
of realizations with $P_{k} \leq P_{\rm max}$
at fixed $L_R=1000~h^{-1}$~Mpc.

Fig.~\ref{WPmax}(a) is obtained  
without trend-subtraction procedure,
i.e. using  Eq.~(\ref{uR}), 
as  the histograms 
in Fig.~\ref{Histo}(a) and (b),
while in Fig.~\ref{WPmax}(b) the 
trend-subtraction procedure 
(see Sect.~\ref{sec:bcg})
is included.  Our additional simulations
show that the curves represented in  
Fig.~\ref{WPmax}(a)
change insignificantly
at filtering out the lowest 
$k \leq 0.04~h$~Mpc$^{-1}$. 

Vertical dashed  lines 
in both panels of Fig.~\ref{WPmax} indicate
significance levels $3\sigma$ (slightly higher)   
corresponding to the  value  $P_{\rm max} \la 10$ 
calculated   following \citet{sc82}.  
Thus in the panel (a) 
the value $[1-W(P_{\rm max})]$
upward each cross of both curves 
($A_m=0.2$  and  $A_m=0$) 
with the vertical dashed line yields a probability
to obtain a peak amplitude  
exceeding the significance level $3\sigma$.\  
We obtain $38\%$ of all realizations 
at $A_m=0$  and  $53\%$ at $A_m=0.2$.
The excess of the probability 
$1-W(P_{\rm max})$ at 
$5 <  P_{\rm max} \la 20$
for the oscillating 
modulation given by  (\ref{f2}) 
relative to the case $A_m=0$  yields an upper 
limit of difference between the two curves. 
Let us note, however,
that for $P_{\rm max} < 10$ and
$P_{\rm max} > 20$ the solid and dashed curves
in Fig.~\ref{WPmax} become indistinguishable.
Thus the impact  of BAO  
on the most significant peaks 
are  small. 

The vertical dot-dashed lines in 
both panels of Fig.~\ref{WPmax} 
also designate
the confidence levels  
$\beta=0.998$ 
but calculated from an equality
$W(P_{\rm max})=99.8\%$ 
applied to the solid curves. 
In addition, one should keep in mind
uncertainties of both the curves which
make them weakly distinguishable and
which we ignore for illustrative purposes.  
 
\section{RADIAL DISTRIBUTION OF BCGs}
\label{sec:bcg}  
For comparison with real cosmological objects we carry out  
calculations of the power 
spectra $P_{\rm R}(k)$  implemented
for the radial  distributions
of brightest 
cluster galaxies (BCGs).  The data on BCGs
are based on the data of SDSS catalogue 
(\citealt{whl12}, \citealt{wh13}). 
We employ 41~420 most luminous ($M_r \leq -23.01$) BCGs
within the spectroscopic redshift range $0.044 \leq z \leq 0.78$
which corresponds to
the comoving-distance range 
$130.8 \leq D_c \leq 1956.8~h^{-1}$~Mpc.
The line-of-sight comoving distances $D_c$ are
calculated according to the  standard expression
(e.g., \citealt{khs97}, \citealt{h99})  
\begin{equation}
D_c(z_l)={c \over H_0}\ \int_0^{z_l} \ 
     {1 \over \sqrt{\Omega_{\rm m}(1+z)^3 + \Omega_\Lambda}}\ {\rm d}z,
\label{Dc}
\end{equation}
where $l=1,2,...$ numerates redshifts $z_l$ of cosmological objects
in the sample, $H_0$ is defined in Introduction, 
$c$ is the speed of light; hereafter
we use the standard $\Lambda$CDM model with 
$\Omega_{\rm m}=0.25$ and 
$\Omega_{\Lambda}=1-\Omega_{\rm m}=0.75$.

We produce a number of radial distributions 
of the BCGs relative
to various reference  points (centres),
with fixed sample length $L_{\rm R} = 1000~h^{-1}$~Mpc
and using  independent bins  $\Delta_c=10~h^{-1}$~Mpc.
For this aim we introduce the Cartesian CS  
employing the data 
(redshifts $z$ and Equatorial coordinates: 
right ascension $-$  $\alpha$ and  declination $-$ $\delta$)    
available on the website 
\footnote{http://zmtt.bao.ac.cn/galaxy\_clusters}.  
 
The transformation to the Cartesian coordinates is 
produced using formulae:
\begin{eqnarray}
& &   X=D_c(z) \cdot \sin(90^\circ - \delta)\ \cos \alpha  
\nonumber                       \\
& &   Y=D_c(z) \cdot \sin(90^\circ - \delta)\ \sin \alpha 
\nonumber                       \\
& &   Z=D_c(z) \cdot \cos(90^\circ - \delta).  
\label{XYZ}
\end{eqnarray}
Accordingly, we introduce the distance 
$r_l$ between a  point object $(X_l, Y_l, Z_l)$  and
a centre of radial distribution $(X_0, Y_0, Z_0)$. 
It allows us to calculate 
[employing Eqs.~(\ref{uR}) and (\ref{PRk})]
$u_R (r)$ and  $P_R (k)$ 
for any sample  of  BCGs. 

Following two main sky domains of the SDSS data
the whole sample of the most luminous BCGs
can be subdivided into two subsamples observed
in the so-called Northern Galactic Cap 
(NGC),\ at $X < 0$\,  (34~541~BCGs) and in the
Southern Galactic Cap 
(SGC),\ at $X > 0$\,  (6~879~BCGs).    
Centres of the radial BCG distributions 
are scattered randomly
over two rectangular prisms situated 
in both galactic hemispheres.
For instance, in the case of NGC the boundaries 
of the prism are
(i) $X < 0$; \,  $-1000 \leq X_0 \leq -400~h^{-1}$~Mpc,\
    $-500 \leq Y_0 \leq +500~h^{-1}$~Mpc,\
    $+100 \leq Z_0 \leq +700~h^{-1}$~Mpc.

\begin{figure*}[htb!]     
\begin{center}
\includegraphics[width=0.95\textwidth]{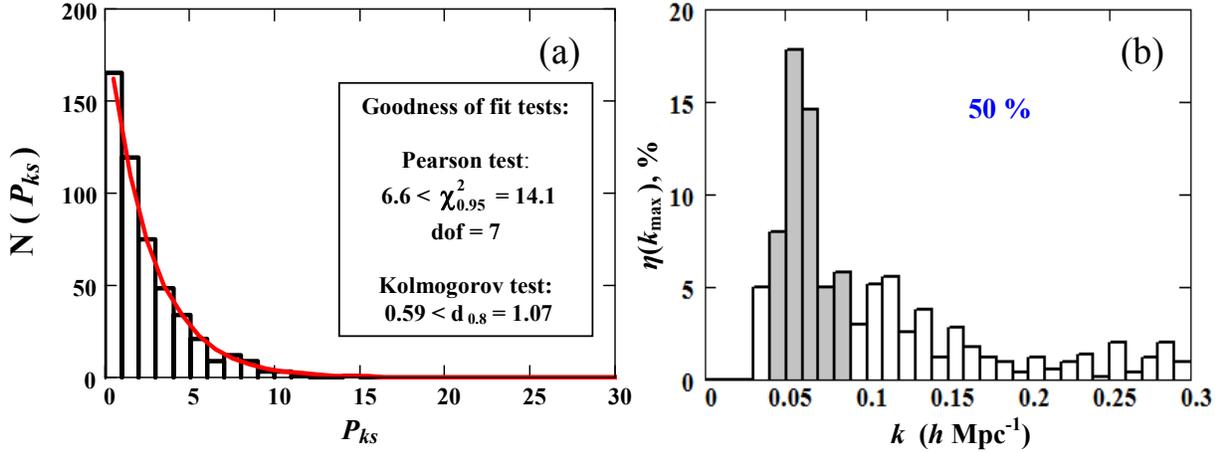}
\caption{
(Colour online)
Statistical properties
of radial power spectra calculated
with using  \protect{(\ref{PRk})}
for the same sample of  34~541
BCGs   as in Fig.~\ref{WPmax}~(b). 
{\it Panel (a)} is a  histogram of 
peak amplitudes $P_{ks}$ at fixed 
$k=k_s=0.0628~h$~Mpc$^{-1}$ 
and  theoretical distribution N$_{th}(P_{ks})$ 
({\it solid line}) 
to find a peak amplitude
within an interval
$n \leq P_{ks} \leq n+1$\, 
($n=0, 1, 2, ..., n_{\rm max}$),
it is similar to Fig.~\ref{3sigma}~(a);\,  
the insert shows results of Pearson's $\chi$-square 
and Kolmogorov's goodness-of-fit tests;\,
{\it panel (b)} $-$ histogram similar  to
Fig.~\ref{Histo}~(c) and (d). 
}  
\label{BCG}
\end{center}
\end{figure*}

For either of the two subsamples we produce 500 
radial distributions $u_R (r)$ 
and calculate the appropriate 
radial power spectra $P_R (k)$. 
Then the averaging over all $P_R (k)$ yields the mean
power spectrum $\langle P_R (k) \rangle$.
In these calculations we use the trend subtraction
procedure described in Sects.~\ref{sec:ms}
to calculate a trend function $N_{\rm tr} (r_b)$ for each 
radial distribution. 
The results of the statistical analysis   
for the NGC subsample
are represented in 
Figs.~\ref{WPmax}(b) 
and  \ref{BCG}.

The thick solid curve $W(P_{\rm max})$ 
in Fig.~\ref{WPmax}(b)
is calculated with the use of Eq.~(\ref{WP}) 
for the sample of 34~541 BCGs.  
The dashed curve in the same panel 
represents  $W(P_{\rm max})$  obtained
in a similar way as
the dashed curve 
in the panel (a) (at $A_2=A_m=0$)  
but making use
a reduced normalizing constant
$A_1' = 340$   in Eq.~(\ref{fCDM}), 
instead of $A_1=470$.  That  
provides a better fit of the
$f_1(k)$ modulating function to an averaged 
radial power spectrum  
$\langle P_{\rm R}(k) \rangle$ produced for the NGC 
sample. Notice that the two
curves (solid and dashed)  
are close. 
In particular, it is seen in 
the panel (b) that the probability
$1-W(P_{\rm max})$ 
for the solid curve
to reveal an excess  of the peak  amplitudes $P_{k}$ over 
the level $3\sigma$ is  $\simeq 28\%$.  
This is close to  $26 \%$ for model dashed 
curve in the same panel. 
In general, we conclude that
the damped oscillations  imitating
the BAO seems to weakly
affect (at any $P_{\rm max}$) the radial 
power spectra of BCGs.   

We verified that 
the distribution of random peak amplitudes
in partial power spectra 
calculated for either of two BCG subsamples
obeys the exponential law,
as in the model simulations of Sect.~\ref{sec:ms},
with the parameter $\lambda (k)$ determined  by the mean
power spectrum $\langle P_{\rm R} (k) \rangle$ at fixed $k$,  i.e.,
$\lambda (k) = \langle P_{\rm R} (k) \rangle^{-1} = f_1^{-1}$.
Fig.~\ref{BCG}(a)  represents 
a histogram of  peak values $P_{ks}$ at the same  
$k=k_s=0.0628$ as in Fig.~\ref{3sigma}(a) and
calculated  for the same
radial distributions of 34~541 BCGs 
as in Fig.~\ref{WPmax}(b). 
The solid line 
demonstrates theoretical number N($P_{ks}$) 
of peak amplitudes  
falling into a bin 
$n \leq P_{ks} \leq n+1$, \, 
$n=0, 1, 2, ..., n_{\rm max}$. 
The curve  N($P_{ks}$) 
is obtained using Eq.~(\ref{calFcalF})
with $\lambda = 1/2.32=0.43$;\,  
$f_1(k_s)= 2.32$ at $A_1'=340$. 

As in Fig.~\ref{3sigma}(a)
the insert also 
gives results of the same  one-side 
goodness-of-fit tests  
between the smooth
theoretical curve (\ref{calFcalF})
and the histogram. 
Both the tests 
demonstrate good agreements 
at the confidence level 0.95 for $\chi$-square test
and 0.8 for the Kolmogorov's test, 
number of bins
in the former case is $n_{\rm max}=10$ and 
consequently dof $=10-3=7$ (stringent criterion), 
while in the latter case the sample 
size is $n_{\rm max}=15$. 
 
The  panel (b) demonstrates the
frequency of occurrence  $\eta(k_{\rm max})$ produced
in similar way  as in  the  panels (c) and (d) 
in Fig.~\ref{Histo}.
One can see that 50$\%$ of the   
dominate peaks $k_{\rm max}$
occupy  the interval $0.04 \leq k \leq 0.09$
of our interest.  This is consistent with
the model histogram  at $A_m=0$  
in Fig.~\ref{Histo}(c) 
with approximately the same probability
within the same interval. It also confirms
the conclusion about weak influence of
the BAO modulation on space distributions
of the BCGs.

\section{CONCLUSION AND DISCUSSION}
\label{sec:cd}
The attention of the present
paper is payed to the 
statistical properties of the radial distributions
examined as spherically uniform shell-like formations
characterized 
only by  a distance from reference 
points (centres). 
We consider quasi-periodicities  
like weak spherical ripples imposed on the 
matter distribution. 
The ripples are quite different for different
centres (observers) and may be smoothed out
by averaging over variety
of the radial distributions. 

The main results of the present consideration 
can be summarized as follows.

(1) Properties of the observational 3D-power spectrum 
of galactic density  fluctuations are considered 
as an origin of quasi-periodicities appearing
in the radial  distributions of cosmologically remote
objects.  
As a simple test of such a consideration
we have implemented simulations 
of modulated Gaussian fields in the $k$-space.   
Modulating function for the Gaussian field has been designed
as $f_{\rm mod} (k) = f_1(k) + f_2(k)$
to reproduce a smoothed part 
of the 3D-power spectrum $-$ $f_1 (k)$,
and to imitate qualitatively 
the  damped oscillations imprinted
in the 3D-power spectrum $-$ $f_2 (k)$.
The latter stands for the baryon 
acoustic oscillations (BAO).

(2) As a result  of modulation of the Gaussian field  
in the  Fourier space  we obtain  
radial  distributions of cosmological objects 
in the  real space, which  
may comprise quasi-periodical
components displaying themselves as peaks 
in the radial power spectra 
and as oscillations 
of the radial correlation functions
(which we do not discuss  here).
The dominate quasi-periods
$\Delta D_c  = 2\pi / k_{\rm max}$ 
variate mainly around $2\pi / k_s$,
where $k_s$ is a maximum 
of the first wave (bump)  of the 
modulating function $f_2 (k)$.

(3) It is shown that
the mean radial power spectrum 
$\langle P_R (k) \rangle$\,  
averaged over variety of partial radial power spectra, 
calculated  relatively 
various centres,  
turns out to be  
rather close (in the limit $-$ equal) 
to the  spherically averaged 
3D power  spectrum (\ref{P3D})
produced for the same modulated random field  
(\ref{U}). Moreover the modulating function $f_{\rm mod}(k)$ 
can be formally used as the 3D power spectrum of  
Gaussian fluctuations (\ref{maineq}). 

(4) Amplitudes of peaks in the partial radial power 
spectra are mainly regulated by the 
smooth modulating function $f_1 (k)$ and,
in a less degree,
by a relative amplitude of the first 
modulating oscillation  
$A_m(k)=f_2(k)/f_1(k)$ around $k=k_s$.
Random  peak heights  at fixed $k$  
satisfy  the exponential distribution with 
the exponential parameter $\lambda = \lambda (k)$
determined by the reciprocal modulating function 
$f_{\rm mod}(k)$. 
It gives quite simple way to estimate 
the significance of the peak amplitudes in the 
radial power spectra. In our result
the significance of the dominant peaks at
$k_{\rm max} \sim k_s$
turns out to be quite
moderate ($ \la 3\sigma$) even for relatively high
peak amplitudes  $10 \la P_{\rm max} \la 20$. 
That means stochastic nature of the 
appropriate quasi-periodicities.

(5) Probability to obtain 
the quasi-periodicity 
in the radial distributions 
with the spectral peak amplitudes belonging
to the interval  
$10 \la P_{\rm max} \la 20$
definitely differs for the cases
$A_m(k_s)=0.2$ (overestimation of the oscillating component) 
and $A_m(k_s)=0$
(elimination of this component),
but almost indistinguishable and
small (a few tenths of a percent)   
for $P_{\rm max} \ga 20$.
It means that modulation
of the Gaussian field
by the oscillatory function $f_2(k)$
imitating the BAO 
hardly affects 
the significance beyond the level 
$\ga 3\sigma$.

(6)  Results of our simulations
of the radial distributions 
specified by 
modulated Gaussian fields  
turn out to be consistent with 
the results of analysis implemented 
for radial distributions of 
the brightest cluster galaxies (BCGs).
The radial power spectra    
calculated for  observational  data on BCGs
and for modulated Gaussian simulations 
at $A_m(k_s)=0$ (i.e.  disregarding
the BAO effects)
demonstrate mutual
consistence of these statistics.  
This consistence can give an
evidence (at least qualitatively)  
in favour of the model of simple modulation
exploited here.

In general, the radial power spectra may display  
rather strong  peaks but estimations
of their significance should be implemented by a
nonstandard  way guided, for instance, by the approach  
suggested in the present paper.  
 
We carried out additionally another way of 
statistical averaging 
of radial distributions 
produced as numerous 
realizations of the modulated random field  
$U({\bf r})$
but simulated relatively to a single centre 
({\it ensemble average}). 
Results turned out to be equal to those, 
described in the present paper,
which is obtained with the use of averaging 
over various centres of the radial distributions
but produced for a single realization of $U({\bf r})$
({\it volume average}).  
We have made sure by simulations 
that the two considered ways of 
averaging are equivalent, that is
our model simulations
do not contradict to the ergodic principle 
for  random fields  (e.g., \citealt{p03}).


Finally, let us remark  that the present work
was inspired by a set of papers,
e.g., by \citet{kp91}, \citet{Dekel92}, \citet{ehss98}, \citet{yoshida01}   
dedicated to  critical examinations 
of the  pencil-beam quasi-periodicity 
at a scale $\sim$ 130~$h^{-1}$~Mpc
found in the pencil-beam surveys 
near the both Galactic poles
by  \citet{beks90},  
and confirmed in further studies by \citet{szetal91,szetal93}. 
Leaving apart  the keen question of the significance assessments,  
the pencil-beam quasi-periodicity 
can not be referred in a straightforward way
to the quasi-periodicity of the
radial distributions discussed here. 
Thus the estimations  
of the peak significance 
produced  for the 
distribution of galaxies along pencil beams 
(inside narrow observational tubes;\,  
e.g., \citealt{kp91}) 
noticeably  differ from the  
estimations suggested here for
the radial galaxy distributions. 
However, both types of spatial oscillations
despite  differences of their scales, 
in principle, could be related in an indirect way.
For instance, there could be different ways to trace 
the same part of cosmic structure. 
This set of questions will be touched 
on by our following  paper.



\bibliographystyle{spr-mp-nameyear-cnd}{}



\end{document}